\documentstyle{article}
\input math_macros

\def \bbook{{\elevenit B Decays}, ed. by S. L. Stone (World Scientific,
Singapore, 1992)}

\def \cn{Collaboration}
\def \cp89{{\elevenit CP Violation,} edited by C. Jarlskog (World Scientific,
Singapore, 1989)}

\def \f79{{\elevenit Proceedings of the 1979 International Symposium on Lepton
and
Photon Interactions at High Energies,} Fermilab, August 23-29, 1979, ed. by
T. B. W. Kirk and H. D. I. Abarbanel (Fermi National Accelerator Laboratory,
Batavia, IL, 1979}
\def \hb87{{\elevenit Proceeding of the 1987 International Symposium on Lepton
and
Photon Interactions at High Energies,} Hamburg, 1987, ed. by W. Bartel
and R. R\"uckl (Nucl. Phys. B, Proc. Suppl., vol. 3) (North-Holland,
Amsterdam, 1988)}
\def \ib#1#2#3{{\elevenit ibid.}~{\elevenbf#1}, #2 (#3)}
\def \ibj#1#2#3{~{\elevenbf#1}, #2 (#3)}
\def \ichep72{{\elevenit Proceedings of the XVI International Conference on
High
Energy Physics}, Chicago and Batavia, Illinois, Sept. 6 -- 13, 1972,
edited by J. D. Jackson, A. Roberts, and R. Donaldson (Fermilab, Batavia,
IL, 1972)}

\def \ite{{\elevenit et al.}}
\def \lkl87{{\elevenit Selected Topics in Electroweak Interactions}
(Proceedings of
the Second Lake Louise Institute on New Frontiers in Particle Physics, 15 --
21 February, 1987), edited by J. M. Cameron \ite~(World Scientific, Singapore,
1987)}
\def \ky85{{\elevenit Proceedings of the International Symposium on Lepton and
Photon Interactions at High Energy,} Kyoto, Aug.~19-24, 1985, edited by M.
Konuma and K. Takahashi (Kyoto Univ., Kyoto, 1985)}

\def \nc#1#2#3{Nuovo Cim. {\elevenbf#1}, #2 (#3)}

\def \pl#1#2#3{Phys. Lett. {\elevenbf#1}, #2 (#3)}
\def \plb#1#2#3{Phys. Lett. B {\elevenbf#1}, #2 (#3)}

\def \prd#1#2#3{Phys. Rev. D {\elevenbf#1}, #2 (#3)}
\def \prl#1#2#3{Phys. Rev. Lett. {\elevenbf#1}, #2 (#3)}
\def \prp#1#2#3{Phys. Rep. {\elevenbf#1}, #2 (#3)}

\def \si90{25th International Conference on High Energy Physics, Singapore,
Aug. 2-8, 1990}
\def \slc87{{\elevenit Proceedings of the Salt Lake City Meeting} (Division of
Particles and Fields, American Physical Society, Salt Lake City, Utah, 1987),
ed. by C. DeTar and J. S. Ball (World Scientific, Singapore, 1987)}
\def \slac89{{\elevenit Proceedings of the XIVth International Symposium on
Lepton and Photon Interactions,} Stanford, California, 1989, edited by M.
Riordan (World Scientific, Singapore, 1990)}
\def \smass82{{\elevenit Proceedings of the 1982 DPF Summer Study on Elementary
Particle Physics and Future Facilities}, Snowmass, Colorado, edited by R.
Donaldson, R. Gustafson, and F. Paige (World Scientific, Singapore, 1982)}
\def \smass90{{\elevenit Research Directions for the Decade} (Proceedings of
the
1990 Summer Study on High Energy Physics, June 25--July 13, Snowmass,
Colorado),
edited by E. L. Berger (World Scientific, Singapore, 1992)}
\def \tasi90{{\elevenit Testing the Standard Model} (Proceedings of the 1990
Theoretical Advanced Study Institute in Elementary Particle Physics, Boulder,
Colorado, 3--27 June, 1990), edited by M. Cveti\v{c} and P. Langacker
(World Scientific, Singapore, 1991)}

\def \zpc#1#2#3{Zeit. Phys. C {\elevenbf#1}, #2 (#3)}
\font\tenbf=cmbx10
\font\tenrm=cmr10
\font\tenit=cmti10
\font\elevenbf=cmbx10 scaled\magstep 1
\font\elevenrm=cmr10 scaled\magstep 1
\font\elevenit=cmti10 scaled\magstep 1

\renewenvironment{thebibliography}[1]
 { \elevenrm
   \begin{list}{\arabic{enumi}.}
    {\usecounter{enumi} \setlength{\parsep}{0pt}
     \setlength{\itemsep}{3pt} \settowidth{\labelwidth}{#1.}
     \sloppy
    }}{\end{list}}

\begin{document}
\renewcommand{\thefootnote}{\dag}
\begin{center}{{\tenbf QUARKONIUM AND POWER LAW POTENTIALS\footnote{Presented
by Aaron K. Grant at DPF 92 Meeting, Fermilab, November, 1992.}\\}
\vspace{-1in}
\rightline{EFI 92-59}
\rightline{November 1992}
\bigskip
\vglue 2.0cm
{\tenrm AARON K. GRANT and JONATHAN L. ROSNER \\}
\baselineskip=13pt
{\tenit Enrico Fermi Institute and Department of Physics, University of
Chicago\\}
\baselineskip=12pt
{\tenit Chicago, IL, 60637\\}

\vglue 0.8cm
{\tenrm ABSTRACT}}
\end{center}
\vglue 0.3cm
{\rightskip=3pc
 \leftskip=3pc
 \tenrm\baselineskip=12pt
 \noindent
The spectra and decay rates of $c \bar c$ and $b \bar b$ levels are well
described, for the most part, by a power-law potential of the form
$V(r)=\lambda(r^{\alpha}-1)/\alpha+{\rm const.}$, where $\alpha\simeq 0$. The
results of an up-to-date fit to the data on spin-averaged levels are presented.
Results on electric dipole transitions in systems bound by power law potentials
are also presented, with applications to the bottomonium system.
\vglue 0.6cm}
\baselineskip=14pt
\elevenrm
Charmonium ($c \bar c$) and bottomonium ($b \bar b$) systems provide a rich
source of information on the interquark force at distances ranging from less
than 0.1 fm to greater that 1 fm.  At short distances our theoretical
prejudices favor a potential which should act like a Coulomb potential
$V(r)=\alpha_s (r)/r$, with $\alpha_s$ becoming smaller at shorter distances
owing to the asymptotic freedom of QCD.$^1$  At long distances, there are both
experimental and theoretical reasons$^2$ to believe that the interquark force
in QCD becomes approximately distance-independent, corresponding to a linear
potential $V(r)\sim r$.  The $c\bar c$ and $b\bar b$ systems appear to lie in
an intermediate range, where a power-law potential $V(r)\sim r^\alpha$ provides
a convenient interpolating form$^3$ between the short-distance Coulomb-like and
long-distance linear behavior.

An early fit to quarkonia spectra$^{4}$ found a power $\alpha\simeq 0.1$.
Since then, data on $P$-wave levels have appeared,$^{5,6}$ and information on
leptonic widths has improved.  It is appropriate to update the earlier fit for
a number of reasons.  The power-law method can be of use in estimating
properties of systems containing $b$ and $c$ quarks, and in interpolating
between these cases to get estimates of $b \bar c$ properties. The power law
method also gives an estimate of the mass difference between the $b$ and $c$
quarks, which can be of use in attempts$^{7}$ to extract the
Cabibbo-Kobayashi-Maskawa matrix element $V_{cb}$ from data on semi-leptonic
$b$ decays.  One would also like to see if there is a consistent pattern of
data signaling a departure from a single effective power at short distances, as
one would expect from the short-distance behavior of QCD.  Finally, power law
potentials can be of use in explaining patterns of electric dipole transition
rates in the $b \bar b$ system.

Since the power-law description does not give an adequate description of
spin-dependent effects, we fit spin-averaged levels. For S waves, we use
\begin{equation}
M(S) = [M (^1 S_0) + 3 M(^3S_1)]/4~~~,
\end{equation}
a combination which eliminates hyperfine splittings. For P waves, we use
\begin{equation}
M(P) = \overline{M}(^3P) \equiv [M(^3P_0) + 3M(^3P_1) + 5M(^3P_2)]/9~~~,
\end{equation}
which eliminates spin-orbit and tensor force splittings.  In the latter case we
assume hyperfine effects are small. In the case of $b \bar b$ levels, the
masses of the $^1$S levels are not experimentally known.  However, the
hyperfine splitting can still be estimated$^{3}$ using information from the
leptonic widths.

We obtain theoretical values for the masses and leptonic widths of the levels
using non-relativistic quantum mechanics.  We find the energies and
wavefunctions of the radial Schr\"odinger equation by solving the dimensionless
equation numerically and then rescaling the dimensionless quantities by
appropriate powers of the mass and coupling constant.$^3$  We take
$V(r)=\lambda (r^\alpha-1)/\alpha+C$. Particle masses are then given by
$M=E+2m_Q$, where E is the binding energy and $m_Q$ is the quark mass.

We present results from three fits.  In each case we minimize $\chi^2$, with
the standard deviations for the energies set equal to  10 MeV, and the standard
deviations for the leptonic widths set equal to the experimental errors. In the
first fit, we consider the levels only. In this case, we find that the best fit
is given by a potential $V(r)\sim r^{-0.045}$, and that the quark mass
difference is $m_b-m_c=$ 3.19 GeV. However, the fit gives little preference for
quark masses, and in fact the best fit is given by $m_b$, $m_c \rightarrow
\infty$.  In the second fit, summarized in Table I, we remedy this by including
the leptonic widths, which are given by
the formula$^{8}$
\begin{equation}
\label{width}
\Gamma(Q\bar{Q}\rightarrow e^{+}e^{-})={{16\pi e_{Q}^2 \alpha^2}\over{M^2}}
         |\Psi(0)|^2 \left [1-{{16\alpha_{s}(m_{Q})}\over{3\pi}} \right ],
\end{equation}
and where we use$^{9}$ $\alpha_s(m_b)=0.189\pm 0.008$, $\alpha_s(m_c)=0.29 \pm
0.02$.  In this case, we find
\begin{equation}
m_{b}=5.24\; {\rm GeV}~~~,~~m_{c}=1.86\; {\rm GeV}~~~,~~
m_b - m_c = 3.38~{\rm GeV}
\end{equation}
for the quark masses, while in the potential
$V(r)=\lambda(r^\alpha-1)/\alpha+C$, we have
\begin{equation}
\alpha=-0.14~~~,~~\lambda=0.808~~~,~~C=-1.305\; {\rm GeV}~~~.
\end{equation}
We find that the fitted masses are correct to within 10 MeV or so, but the
leptonic widths are off, typically, by 20 to 30 percent.  In an attempt to
remedy this, in the third fit, we include an {\elevenit ad hoc} relativistic
correction to the leptonic widths. We correct Eq. (\ref{width}) by introducing
a factor of the form $(1+K\langle v^2/c^2\rangle)$, and treat $K$ as a free
parameter. In this case, we find $\alpha=-0.12$, $\lambda=0.801$, and
$C=-0.772$ GeV, while for the quark masses we have $m_b =4.96$ GeV, $m_c=1.56$
GeV, and $m_b - m_c = 3.40$ GeV. The constant $K$ is found to be 1.25.  The
fitted quantities are given in Table I.  The relativistic correction gives only
a marginal improvement in the fit.
{\small
\begin{table}
\caption{J/$\psi$ and $\Upsilon$ masses and leptonic widths}
\label{tab2}
\begin{center}
\begin{tabular}{ccccccc}
\hline\hline
Particle       &         & Mass (GeV)&            &                &
       Width (keV)   &               \\
               & (Expt.) & (NR$^a$)  & (RC$^b$)   & (Expt.)        &
       (NR$^a$)      & (RC$^b$)      \\ \hline
J/$\psi$(1S)   & 3.068   & 3.077     & 3.079      &  5.36$\pm$0.29 &
       6.41$\pm$0.43 & 6.29$\pm$0.43  \\
J/$\psi$(2S)   & 3.663   & 3.654     & 3.654      & 2.14$\pm$0.21  &
       2.03$\pm$0.13 & 2.04$\pm$0.13  \\
J/$\psi$(1P)   & 3.525   & 3.524     & 3.522      &      --        &
             --      &        --      \\
$\Upsilon$(1S) & 9.449   & 9.420     & 9.423      & 1.34$\pm$0.04  &
       1.21$\pm$0.02 & 1.18$\pm$0.02 \\
$\Upsilon$(2S) & 10.018  & 10.044    & 10.042     & 0.563$\pm$0.14 &
     0.477$\pm$0.009 & 0.475$\pm$0.009 \\
$\Upsilon$(3S) & 10.351  & 10.358    & 10.358     & 0.44$\pm$0.07  &
     0.285$\pm$0.006 & 0.284$\pm$0.005 \\
$\Upsilon$(4S) & 10.578  & 10.564    & 10.567     & 0.24$\pm$0.05  &
     0.197$\pm$0.004 & 0.200$\pm$0.004 \\
$\Upsilon$(1P) & 9.900   & 9.903     & 9.900      &       --       &
            --       &       --        \\
$\Upsilon$(2P) & 10.260  & 10.269    & 10.267     &      --        &
            --       &       --        \\
$\Upsilon$(1D) & --      & 10.181    & 10.177     &      --        &
            --       &       --        \\
$\Upsilon$(2D) & --      & 10.436    & 10.435     &      --        &
            --       &       --        \\
\hline \hline

\end{tabular}
\end{center}
\vskip -0.4cm
\centerline{$^a$No relativistic corrections \qquad $^b$With relativistic
corrections}
\vskip -0.2cm
\end{table}
}
In the fits to masses and leptonic widths, we find that $m_b$ and $m_c$ are not
particularly well determined.  However, the mass difference $m_b-m_c$ appears
to be more stable, with a value around 3.39 GeV.

We can use the fitted potentials to estimate the centers of gravity of a few
low-lying levels of $b\bar c$, $b \bar s$ and $c \bar s$ levels. For the $b\bar
c$ system, we find
\begin{equation}
M_{b\bar c}(1S)=6.304~{\rm GeV,}~~M_{b\bar c}(2S)=6.898~{\rm GeV,}~~~
M_{b\bar c}(1P)=6.764~{\rm GeV,}
\end{equation}
while for the $b \bar s$ and $c \bar s$ levels, we have
\begin{equation}
M_{c\bar s}(1S)=2.085~{\rm GeV,}~~M_{c\bar s}(1P)=2.509~{\rm GeV,}~~~
M_{b\bar s}(1S)=5.401~{\rm GeV.}
\end{equation}
The experimental spin-averaged masses$^{10,11}$ of the $c \bar s$ states are
2.075 GeV for the 1S level, and 2.536 GeV for the 1P level.  We have also
estimated the 1S - 2S splitting in toponium.  Taking $m_t=$130 GeV, we find the
splitting to be roughly 0.8 GeV.  We expect that this is a conservative lower
bound, since the short-distance behavior of QCD (which will make this splitting
larger) should be important in this case.

Power law potentials also offer some insight into the strengths of various E1
transitions in the $b \bar b$ system.$^{12}$ Experimentally,$^{10}$ we find
that the 3S-1P transition is suppressed relative to the 3S-2P, and that the
2P-1S transition is suppressed relative to the 2P-2S.  We can gain some insight
into this by considering the the radial dipole matrix elements $\langle
u_{n\ell}|r| u_{n'\ell\pm1} \rangle$, where $n$ denotes the number of nodes in
the radial wavefunction $u_{n\ell}$. We can approximately evaluate these matrix
elements by considering the large $\ell$ limit.  In this case, we expand the
potential in the radial Schr\"odinger equation about the point the point
$\bar{r}=\ell^{2/(2+\alpha)}$.  The potential then has the form of a harmonic
oscillator potential, plus anharmonic terms which we include perturbatively.
For the dipole matrix elements, we find
$$
\langle u_{n\ell}|r|u_{n\ell-1}\rangle=\ell^{2/(\alpha+2)},~~~
\langle u_{n\ell-1}|r|u_{n-1\ell}\rangle=\sqrt{n}\Psi_{+}(\alpha)
                     \ell^{(2-\alpha)/(4+2\alpha)},
$$
\begin{equation}
\label{dipole}
\langle u_{n-1\ell-1}|r|u_{n\ell}\rangle=\sqrt{n}\Psi_{-}(\alpha)
                     \ell^{(2-\alpha)/(4+2\alpha)},~~~
\langle u_{0\ell}|r|u_{2\ell-1}\rangle=\Phi(\alpha) \ell^{-\alpha/(2+\alpha)},
\end{equation}
where $\Psi_{\pm}$ and $\Phi$ are functions which depend only on $\alpha$.
We see from Eqs.~(\ref{dipole}) that transitions with $\Delta n=0$ are
dominant, and that the others are suppressed by a factor $\ell^{-\Delta n/2}$
in the limit of large $\ell$. Furthermore, in the region of interest for
quarkonium, $\alpha\simeq 0$, the coefficient $\Phi$ is quite small:
$\Phi(0)\simeq -0.04$.  This further suppresses the E1 rate for $\Delta n =2$
transitions.  Direct numerical calculations show that this is also the case for
small $\ell$: we find that the 3S-1P rate is dramatically suppressed relative
to the 3S-2P rate. Consequently, it appears that the suppression of E1
transitions with nonzero $\Delta n$ is a general property of power law
potentials.

We can use the ratios of radial dipole matrix elements for certain transitions
in the $b \bar b$ system to place limits on the power $\alpha$.  From the
experimental data$^{6,10}$ we deduce the ratios
\begin{equation}
{\langle 3S|r|1P\rangle\over{\langle 3S|r|2P\rangle}}=0.016\pm0.004,~~~
{}~~~~~~
{\langle 2P|r|1S\rangle\over{\langle 2P|r|2S\rangle}}=0.117\pm0.014.
\end{equation}
Computing these ratios numerically and comparing with experiment, we find that
$\alpha$ is constrained to lie in the region $-0.2<\alpha<0$, consistent with
what was found from the fit to levels and leptonic widths. Finally, we can
estimate $m_b$ using the rates of the 3S-1P, 3S-2P, and 2S-1P electric dipole
transitions.  Extracting the radial dipole matrix elements from the data, we
find that the potential model favors a value $m_b\simeq 4.0\pm0.9$ GeV.  This
is smaller than expected, and in particular we find that the 2S-1P transition
yields the surprisingly small value $m_b=3.45\pm0.44$ GeV.  This is an
indication that either the potential model is inadequate for describing E1
rates, or that the $\Upsilon$(2S) total width has been overestimated.

We thank Eric Rynes for collaboration on the results of Ref. 3.  This work was
supported in part by the United States Department of Energy through Grant No.
DE FG02 90ER 40560.

\end{document}